\documentclass[pra, twocolumn,showpacs,amsmath,amssymb,superscriptaddress]{revtex4-1}

\usepackage{graphicx}% Include figure files
\usepackage{bbold}% bold math

\newcommand{\ket}[1]{\ensuremath{\left|#1\right\rangle}}


\begin{document}

\title{$T_{1}$-Echo Sequence - Protecting the state of a qubit in the presence of coherent interaction}

\author{Clemens M\"uller}
	\affiliation{D\'epartement de Physique, Universit\'e de Sherbrooke, Sherbrooke, Qu\'ebec, Canada J1K 2R1}
	\affiliation{Institut f\"ur Theorie der Kondensierten Materie, Karlsruhe Institute of Technology, D-76128 Karlsruhe, Germany}
	
\author{Alexander Shnirman}
	\affiliation{Institut f\"ur Theorie der Kondensierten Materie, Karlsruhe Institute of Technology, D-76128 Karlsruhe, Germany}
	\affiliation{DFG-Center for Functional Nanostructures (CFN), D-76128 Karlsruhe, Germany}

\author{Martin Weides}
	\affiliation{National Institute of Standards and Technology, Boulder, Colorado 80305, USA}
	\affiliation{Physikalisches Institut, Karlsruhe Institute of Technology, D-76128 Karlsruhe, Germany}
\date{\today}

\begin{abstract}
	We propose a sequence of pulses intended to preserve the state of a qubit in the presence of strong, coherent coupling to another quantum system. 
	The sequence can be understood as a generalized SWAP and works in formal analogy to the well-known spin echo. 
	Since the resulting effective decoherence rate of the qubits state is strongly influenced by the additional system, this sequence might serve to protect its quantum state
	as well as negating the effects of the coherent coupling. 
	A possible area of application are large scale quantum computing architectures, where spectral crowding of the resources might necessitate a method to mitigate residual couplings.
\end{abstract}

\maketitle

	Any future large scale quantum computation architecture will involve a large number of quantum-coherent elements. 
	Any building blocks which employ similar fundamental concepts (i.e. superconducting circuits, quantum dots, ions etc..) will necessarily 
	have level-energies similar in frequency to many others.
	This clustering of energies may lead to unwanted, spurious couplings between different elements, possibly disturbing the operation of the whole.
	Proposed ways around this problem involve tuneable coupling elements with large on-off ratios~\cite{Pinto:2010, Bialczak:2011} 
	as well as hybrid architectures~\cite{Kubo:2011, Kubo:2012, Frey:2012}, where elements of different origin and thus different energy scales are used for different tasks.
	
	In this article we propose a method to dynamically negate the effects of coherent coupling of a qubit to another quantum system. 
	Our proposal consists of a series of pulses applied to the composite system and is inspired by the well known spin-echo sequence~\cite{Hahn:1950}, 
	which is designed to mitigate slow-noise fluctuations in the dynamics of a qubit. 
	Since in the course of the time-evolution the state of the qubit will be partly transferred into the additional quantum system, 
	its properties will strongly influence the resulting state. Specifically, when considering a qubit coupled to a quantum memory element, 
	we find an improvement in the relaxation rate of the qubit. The sequence might therefore also serve to protect the state of a qubit from energy relaxation.
	
	Schemes working towards protecting a qubit state with the use of quantum memory traditionally focus on a full SWAP of the quantum information from qubit to storage element. 
	The information then resides in the memory for the storage time $T$ after which another SWAP sequence is used to write it back into the qubit, 
	see e.g.~\cite{Neeley:2008, Kubo:2011, Zhu:2011}.
	In contrast to this traditional approach, the sequence we propose only partially transfers the state of the qubit into the memory. 
	The possible protection is therefore weaker as in the usual approach but on the other hand application of the necessary pulses can be significantly faster than for a full SWAP.
	
%	The protection achieved by applying the sequence depends on the ratio of qubit and memory energy loss 
%	and is in general half as good as what can be achieved with a full SWAP. 
%	The $T_{1}$-echo can thus bee seen as lying somewhat in between a full SWAP and not applying any protection at all.
%	The time needed for the application of the $T_{1}$-echo sequence is in general shorter and in the worst case equally long as for a traditional SWAP protocol. 
%	The protection is therefore also achievable for shorter storage times.
%	
\section*{Introduction}
	
	\begin{figure}[htbp]
		\begin{center}
			\includegraphics[width=.9\columnwidth]{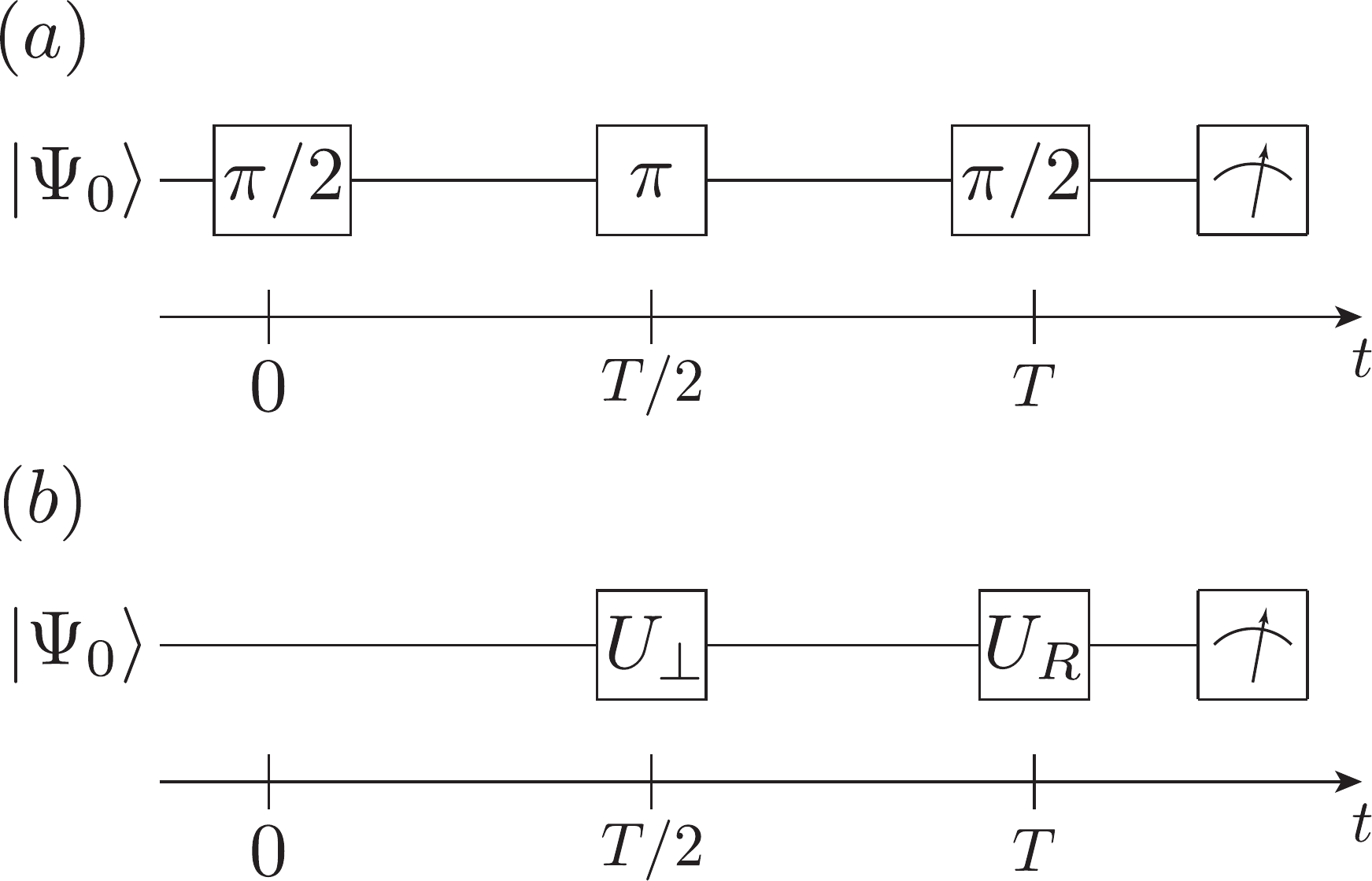}
			\caption{Schematic representation of the pulse sequence for (a) standard spin-echo and (b) the proposed $T_{1}$-echo sequence.
				In both cases, the state of the qubit is read out after time $T$. 
				Also in both cases, the application of echo- and recovery-pulses leads to a final state which is independent of $T$.
				For details, see text.
			}
			\label{fig:PulseSequenceAll}
		\end{center}
	\end{figure}
	
	Our proposal is inspired by the well known spin- or Hahn-Echo sequence~\cite{Hahn:1950}, 
	a schematic depiction of which is shown in Fig.~\ref{fig:PulseSequenceAll} (a).
	The spin-echo sequence is designed to protect a two-level system (TLS) in a Ramsey experiment from the effect of slow, 
	environmentally induced fluctuations of its level-splitting. It finds wide application in NMR and more recently the field of quantum computation~\cite{Ithier:2005}.
	In a typical Ramsey experiment, a slightly detuned $\pi/2$-pulse is applied to a two-level system. 
	After some free evolution time $T$ the same pulse is applied again and the population of the system is measured. 
	The signal as a function of time will show the beating of the state-vector against the rotating frame 
	as defined by the pulse, at a frequency given by the pulse detuning $\delta\epsilon$. 
	If a low frequency noise source acts on the value of the level splitting $\epsilon$ of the two-level system, this will similarly affect the detuning $\delta\epsilon$, and thus 
	subsequent runs of the experiment will lead to different results. This phenomenon is known as pure dephasing.
	The spin-echo sequence now consists of the application of an additional $\pi$-pulse, the echo pulse, after half the evolution time $T/2$. 
	Due to the effect of this pulse, the state vector will, during the second half of the time-evolution, effectively retrace the path it took in the first half.
	The system will at time $T$ thus always reach its initial state again and the effects of fluctuations at frequencies, which are slower than the inverse evolution-time $1/T$, 
	are cancelled by the sequence.

	For what we call the $T_{1}$-echo sequence we want to apply the same principles to the case of two coupled quantum systems, namely a qubit and a quantum memory. 
	In the case of strong coupling between the two, and for an initial state where the qubit is excited while the memory rests in its groundstate, 
	the system will undergo coherent oscillations where the excitation is shifted periodically between the two parts. 
	The frequency of these oscillations depends on the strength of the coherent coupling as well as on the detuning between the two systems. 
	Assuming both these quantities to be constant in time, application of an echo pulse after half the evolution time $T/2$ will cancel the effect
	of the coherent oscillations. Since we have a coupled system to which we apply this pulse sequence, the resulting effective decay of the qubit state will 
	be strongly influenced by the decay of the memory~\cite{Muller:2009}. 
	In the case where the additional quantum system shows better coherence properties than the qubit, this will allow us to find an effective protection of the qubit state from energy relaxation.

	Fig.~\ref{fig:PulseSequenceAll} (b) gives a depiction of the proposed pulse sequence and compares it to the case of a spin-echo protocol.
	In the $T_{1}$-echo sequence, the system evolves freely for a time $T/2$. After this period of free evolution, an echo pulse $\hat U_{\perp}$ is applied,
	which corresponds to a $\pi$-rotation around an axis perpendicular to the original axis of rotation. After another period $T/2$ of free evolution,
	the state reached by the system will no longer depend on the total time $T$ it took to reach it. This state will in general, however, not be the starting state
	and thus has to be corrected by the application of a recovery pulse $\hat U_{R}$.
	The exact nature and possible realizations of the pulses $\hat U_{\perp}$ and $\hat U_{R}$ will be explained in detail below.

\section*{System}

	We focus our theoretical analysis on a system consisting of two coherently coupled two-level systems, one being the qubit 
	and the other one we will call a memory element from now on.
	The Hamiltonian reads
	\begin{equation}
		\hat H_{0} = -\frac{1}{2} \epsilon_{q} \sigma_{z} - \frac{1}{2} \epsilon_{m} \tau_{z} + \frac{1}{4} v_{\perp} \left( \sigma_{+} \tau_{-} + \sigma_{-} \tau_{+} \right) \,,
		\label{eq:H0}
	\end{equation}
	where the $\sigma$ and $\tau$ are the Pauli-matrices describing qubit and memory, respectively. 
	The qubit has a level-splitting of size $\epsilon_{q}$ while the memory is at energy $\epsilon_{m}$ 
	and they are coupled with a coupling strength $v_{\perp}$. 
	We write the coupling in the rotating wave approximation, assuming $\epsilon_{q}, \epsilon_{m} \gg v_{\perp}$.
	We also consider the regime where qubit and memory are close to resonant, $\epsilon_{q} \approx \epsilon_{m}$, 
	meaning the coupling between the two is effective, i.e. important for the time-evolution.
	
	In the validity range of the rotating wave approximation, the above Hamiltonian preserves the total number of excitations. 
	This property, when neglecting decoherence, enables us to effectively decouple the dynamics for different numbers of excitation.
	For the application of our sequence, we assume the memory to be initially in its groundstate $\ket{0_{m}}$. 
	In this case, and under the action of the Hamiltonian Eq.~\eqref{eq:H0}, in order to show protection of the qubit state it is sufficient to show
	that the total groundstate of the full system $\ket{0_{q}, 0_{m}}$ is unaffected and that we can bring the state $\ket{1_{q}, 0_{m}}$ back to itself.
	We will not be concerned with dynamic or geometric phases acquired during time evolution since they can be negated using single qubit gates.
	
	The restriction on a two-level system as the memory used is not necessary for the implementation of the $T_{1}$-Echo protocol, 
	we use Eq.~\eqref{eq:H0} as the simplest conceivable model. 
	The original motivation for these work came from work with superconducting qubits coupled to intrinsic two-level defects~\cite{Neeley:2008, Lisenfeld:2010, Mariantoni:2011}.
	This system naturally realizes the Hamiltonian Eq.~\eqref{eq:H0} and, due to the defects possible long coherence times~\cite{Lisenfeld:2010}, 
	might serve as a testing ground for our proposal.
	Another possibility for use as a memory would be a harmonic oscillator in the quantum regime as was realized e.g., in Refs.~\cite{Hofheinz:2008, Mariantoni:2011}. 
	As long as no more than one excitation (i.e. either one photon in the resonator or the qubit in its excited state) in total is involved in the dynamics, 
	the underlying Jaynes-Cummings Hamiltonian will lead to the same results as the ones presented here.
	Recent experimental efforts have succeeded in coherently transferring quantum information from a qubit 
	to an ensemble of spins in the solid state~\cite{Kubo:2011, Zhu:2011, Cai:2012}, 
	giving yet another possible experimental realization of the model Eq.~\eqref{eq:H0}. 
	In these ensemble systems, the weak individual couplings $g$ of the single spins act collectively to realize a strong effective coupling $\propto \sqrt{N} g$ 
	to a collective state of the ensemble. The effective Hamiltonian is again of the Jaynes-Cummings type and thus can be reduced to Eq.~\eqref{eq:H0} for a single excitation.
	We want to note that there are other transfer schemes proposed in the literature which, rather than use a direct existing interaction between two physical systems, 
	induce such an interaction by use of e.g. laser or microwave excitation, see e.g.~\cite{Morton:2008}. 
	For such situations, our scheme seems impractical to realize since the interaction term in Eq.~\eqref{eq:H0} will be necessary at all times for our proposal.

\section*{Evolution on effective Bloch-Sphere}

	\begin{figure}[htbp]
		\begin{center}
			\includegraphics[width=\columnwidth]{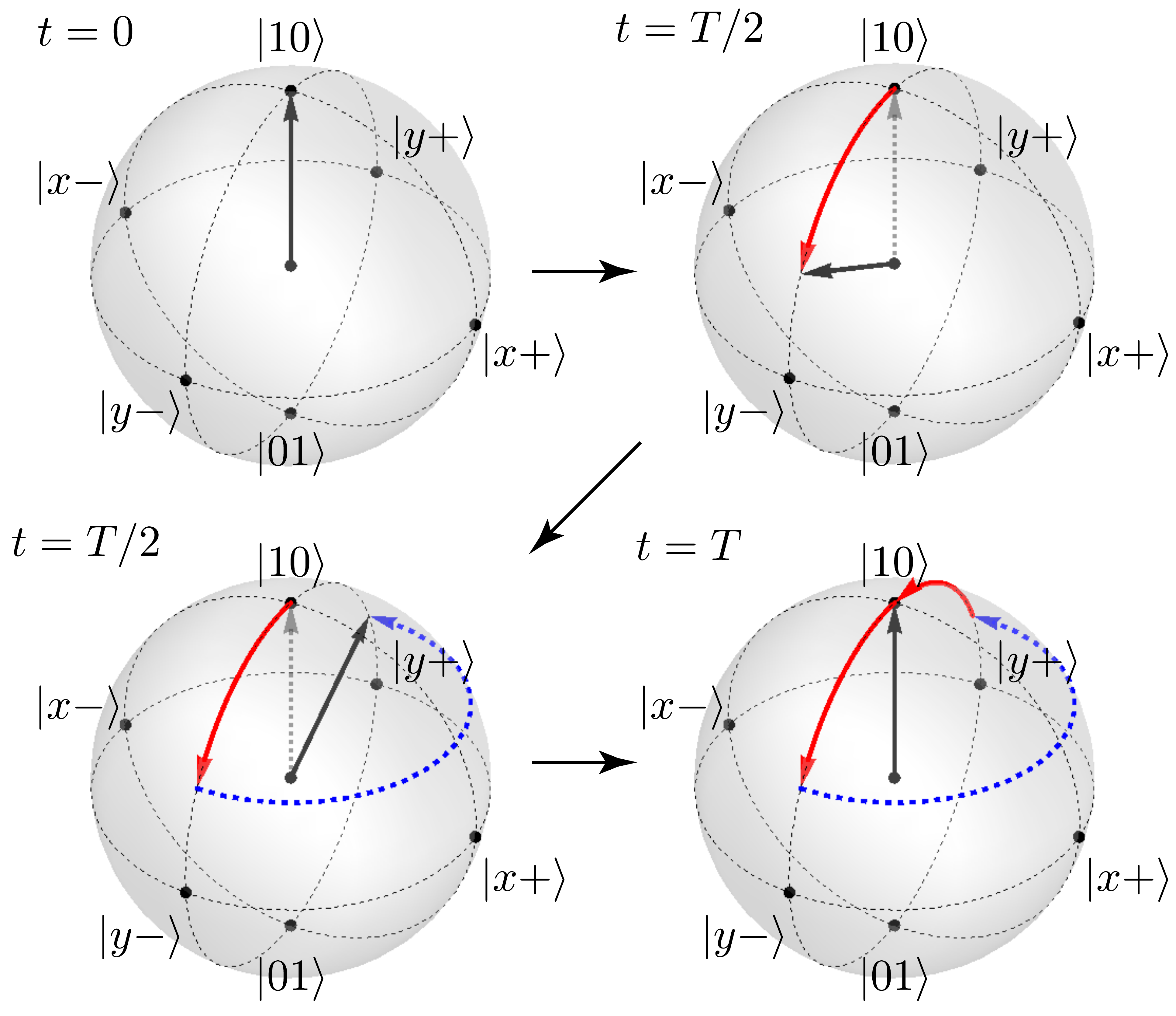}
			\caption{(Color online) 
				Step by step illustration of the $T_{1}$-echo sequence on the effective one-excitation Bloch-sphere for the resonance case, $\delta\omega=0$, 
				with a starting state of $\ket{1_{q} 0_{m}}$, for details see text. 
				Red arrows show the path of the state vector during the two timespans of free evolution, the blue dashed arrow represents the path during the 
				echo pulse $\hat U_{\perp}$. The recovery pulse is not shown, since for $\delta\omega=0$ we have $\ket{\Psi_{T}} = -i  \ket{\Psi_{0}}$.
				The illustration depicts a free evolution time of $T = 2\pi / ( 3 \omega_{0} ) $.
				In this illustration we assume infinitely sharp pulses, i.e. no time is needed for the application of $U_{\perp}$. 
			}
			\label{fig:BlochSphereInit}
		\end{center}
	\end{figure}
	
	\begin{figure}[htbp]
		\begin{center}
			\includegraphics[width=.65\columnwidth]{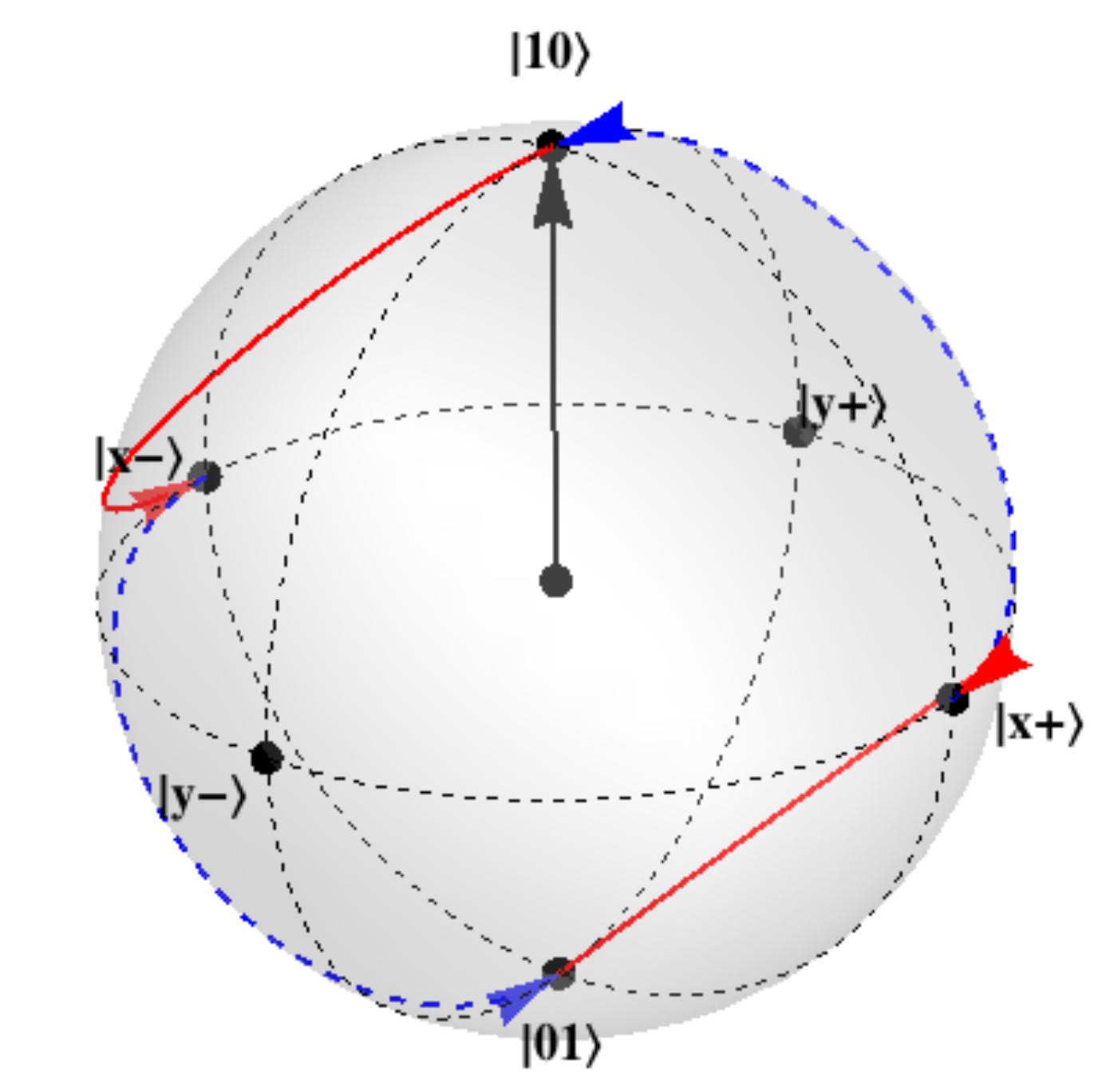}
			\caption{(Color online) Illustration of a full $T_{1}$-echo sequence on the effective Bloch-sphere. Red arrows show the path of the state vector during the 
				free evolution periods, while the dashed blue arrows show the effects of the echo pulse $\hat U_{\perp}$ and the recovery pulse $\hat U_{R}$.
				The black arrow depicts the state of the system at start and end of the sequence.
				The depicted situation corresponds to $\delta\omega = -v_{\perp}$, i.e., $\xi_{0} = 3\pi / 4 $, where we find $\ket{\Psi_{T}} = \ket{x+}$ 
				and a total evolution time of $T = 2\pi / \omega_{0}$.
			}
			\label{fig:BlochSphereEnd}
		\end{center}
	\end{figure}
	
	\begin{figure}[htbp]
		\begin{center}
			\includegraphics[width=.9\columnwidth]{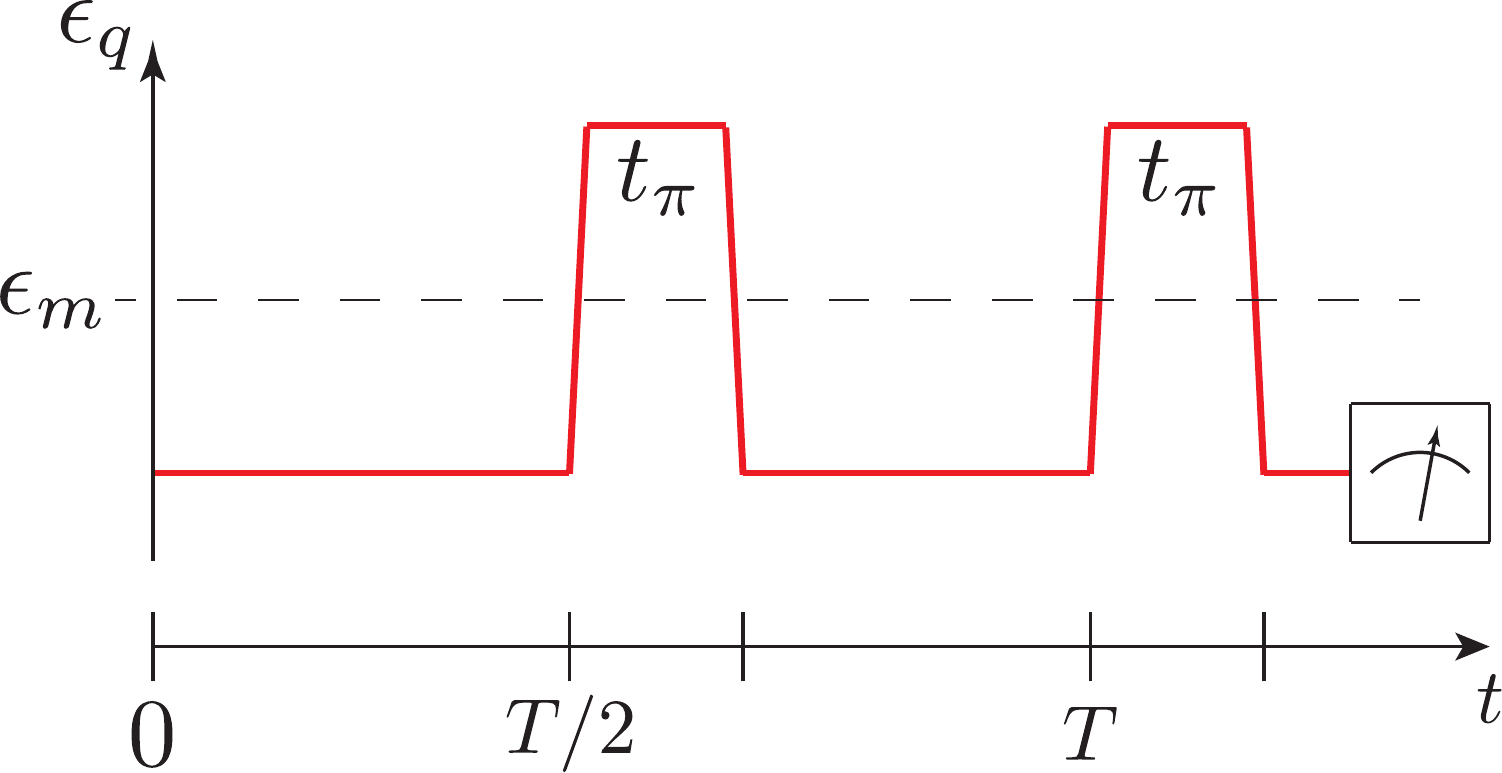}
			\caption{(Color online) Illustration of the $T_{1}$-echo sequence using only detuned Hamiltonian evolution. 
				Both pulses, $\hat U_{\perp}$ and $\hat U_{R}$ can be replaced by a period of detuned free evolution at detuning 
				$\delta\omega_{1} = - v_{\perp}^{2} / \delta\omega$ for the time $t_{\pi}$. 
				The red line depicts the qubit's level splitting at different times. The resonance condition with the memory, $\epsilon_{q} = \epsilon_{m}$, 
				is indicated by the dashed line in the middle.
				The situation depicted corresponds to the case of initially $\delta\omega = v_{\perp}$ where $\delta\omega_{1} = -\delta\omega$.
			}
			\label{fig:PulseSequenceDetuned}
		\end{center}
	\end{figure}
	
	To illustrate the workings of the $T_{1}$-Echo sequence, we will first focus on the case of purely unitary evolution, i.e. zero decoherence. 
	In this case, and starting with the qubit in its excited state and the memory in its groundstate, $\ket{1_{q}, 0_{m}}$, the time evolution is restricted to a subspace of the full
	Hilbert-space containing only one excitation. 
	We can describe this subspace by an effective Bloch-sphere, spanned by the states containing only a single excitation in either the qubit or the memory,
	i.e. $\mbox{span}\{ \ket{1_{q}, 0_{m}}, \ket{ 0_{q}, 1_{m} } \}$.
	We will first give the expressions  for $\hat U_{\perp}$ and $\hat U_{R}$ and explain their effects on the time evolution.
	Later in this chapter we will also show how to realize this pulses without the need for high-frequency controls, using only detuned Hamiltonian evolution governed by Eq.~\eqref{eq:H0}.
	
	The states describing the effective Bloch-Sphere are
	\begin{align}
		\ket{z+} &= \ket{1_{q}, 0_{m}} \equiv (0,0,1) \,,\nonumber\\
		\ket{z-} &= \ket{0_{q}, 1_{m}} \equiv (0,0,-1)  \,,\nonumber\\
		\ket{x+} &= \frac{1}{\sqrt{2}} \left( \ket{1_{q}, 0_{m}} + \ket{0_{q}, 1_{m}} \right) \equiv (1,0,0) \,,\nonumber\\
		\ket{x-} &= \frac{1}{\sqrt{2}} \left( \ket{1_{q}, 0_{m}} - \ket{0_{q}, 1_{m}} \right) \equiv (-1,0,0) \,,\nonumber\\
		\ket{y+} &= \frac{1}{\sqrt{2}} \left( \ket{1_{q}, 0_{m}} + i \ket{0_{q}, 1_{m}} \right) \equiv (0,1,0)  \,,\nonumber\\
		\ket{y-} &= \frac{1}{\sqrt{2}} \left( \ket{1_{q}, 0_{m}} - i \ket{0_{q}, 1_{m}} \right)  \equiv (0,-1,0) \,,
	\end{align} 
	where the indices $q, m$ denote states of the qubit and the memory respectively and the vectors $(x,y,z)$ describe points on the pseudo-3D Bloch-sphere.
	
	Restricting the Hilbert space to the one-excitation subspace, the time evolution operator corresponding to Eq.~\eqref{eq:H0} can be written as 
	\begin{align}
		\hat U_{1}(t, \xi_{0}) &= \exp{\{- i \hat H_{1} t\}} \nonumber \\
			&= \cos{(\frac{1}{2} \omega_{0} t)} \mathbb{1}_{2} - i \sin{(\frac{1}{2} \omega_{0} t)} \left( 
				\begin{array}{cc}
					 - \cos{\xi_{0}} & \sin{\xi_{0}} \\
					\phantom{-}\sin{\xi_{0}} & \cos{\xi_{0}}
			\end{array} \right) \,,
		\label{eq:U1}
	\end{align}
	with $\omega_{0} = \sqrt{v_{\perp}^{2} + \delta\omega^{2}}$, $\tan{\xi_{0}} = v_{\perp} / \delta\omega$ and we choose for the detuning angle $\xi_{0} \in [ 0,\pi ]$.
	Here, $\delta\omega = \epsilon_{q} - \epsilon_{m}$ is the detuning between qubit and memory and we write $\hat H_{1}$ for the part of the Hamiltonian $\hat H_{0}$
	which is acting on the one-excitation subspace.
	The expression Eq.~\eqref{eq:U1} can be interpreted as free precession on the Bloch-sphere around the axis $\vec n_{0} = (\sin{\xi_{0}}, 0, \cos{\xi_{0}}) $ 
	with frequency $\omega_{0}$. The rotation axis $\vec n_{0}$ forms an angle $\xi_{0}$ with the $z$-axis. 
	For zero detuning, $\delta\omega = 0$, we find $\xi_{0} = \pi/2$ and therefore $\vec n_{0} = (1, 0, 0)$, i.e. the precession takes place around the x-axis. 
	In the case of strong detuning, $\delta\omega \gg v_{\perp}$, we have $\xi_{0} = 0$ and thus the rotation axis is $\vec n_{0} = (0, 0, 1)$. 
	If the initial state is also lying on the $z$-axis, i.e., $\ket{1_{q}, 0_{m}}$ no rotation will take place. 
	This is equivalent to saying that for sufficiently strong detuning between the two systems, their coupling is no longer affecting the time-evolution.
	Finally, for non-zero but finite detuning, the resulting rotational axis lies  somewhere in the middle between the x- and z-axis. 
	E.g., for $\delta\omega = v_{\perp}$, we find $\xi_{0} = \pi / 4$ and $\vec n_{0} = \frac{1}{\sqrt{2}}(1, 0, 1)$.
	For our choice of interaction term in the Hamiltonian Eq.~\eqref{eq:H0}, the axis around which rotations take place is thus fixed on the $xz$-plane of the effective Bloch-sphere. 
	It is important to note that this is no special property of our choice of interaction. 
	In fact any physical interaction corresponding to an exchange of energy between the qubit and the memory will give a similar restriction of the axis of rotation 
	to a plane in the effective Bloch-sphere picture.
	
	In keeping with the ideas of spin-echo, we want to apply a $\pi$-pulse around an axis which is perpendicular to the axis $\vec n_{0}$ of free evolution, 
	and we want to apply this pulse after half the total evolution time $T$.
	We have some freedom in the choice of this axis, and will choose it lying again in the $xz$-plane of the one-excitation Bloch-sphere. 
	This selection will later enable us to perform this rotation using only detuned Hamiltonian evolution, without the need for high frequency controls.
	The axis $\vec n_{\perp}$ lying on the $xz$-plane, and perpendicular to $\vec n_{0}$ is simply $\vec n_{\perp} = (- \cos{\xi_{0}}, 0, \sin{\xi_{0}})$.
	An operator corresponding to a $\pi$-pulse around $\vec n_{\perp}$ is given by
	\begin{equation}
		\hat U_{\perp}(\xi_{0}) = i \left( \begin{array}{cc}
				\sin{\xi_{0}} & \phantom{-} \cos{\xi_{0}} \\
				\cos{\xi_{0}} & - \sin{\xi_{0}}
			\end{array} \right) \,.
		\label{eq:UEcho}
	\end{equation}
	After application of the echo pulse $\hat U_{\perp}$, another period of free evolution for the time $T/2$ brings the system into the state $\ket{\Psi_{T}}$, 
	which is independent of the total evolution time $T$.
	The state $\ket{\Psi_{T}}$ before the application of the recovery pulse $\hat U_{R}$ is given by
	\begin{align}
		\ket{\Psi_{T} } &= \hat U_{1}(T/2, \xi_{0}) \: \hat U_{\perp}(\xi_{0}) \: \hat U_{1}(T/2, \xi_{0}) \: \ket{1_{q}, 0_{m}} \nonumber \\
			&= i \left( -\sin{\xi_{0}} \ket{1_{q}, 0_{m}} + \cos{\xi_{0}} \ket{0_{q}, 1_{m}} \right) \nonumber\\
			&\equiv (-\sin{2\xi_{0}}, 0, -\cos{2\xi_{0}}) \,,
		\label{eq:PsiT}
	\end{align}
	meaning it returns again to the $xz$-plane of the effective Bloch-sphere. It is important to note again that the state $\ket{\Psi_{T}}$ is independent of the 
	evolution time $T$, i.e., the echo pulse works as expected.
	
	Fig.~\ref{fig:BlochSphereInit} shows an illustration of the full time-evolution without application of the final recovery pulse $\hat U_{R}$.
	The black arrow in the pictures shows the state of the system at a given time $t$ when starting from state $\ket{1_{q}, 0_{m}}$
	while the red arrows depict the path the state vector took on the Bloch-sphere
	due to free Hamiltonian evolution in the initial and final timespans of length $T/2$. 
	The blue dashed arrow finally depicts the path taken during application of the echo-pulse $\hat U_{\perp}$.
	The situation shown in Fig.~\ref{fig:BlochSphereInit} is for zero detuning between qubit and memory, $\xi_{0} = \pi/2$, where $\ket{\Psi_{T}} = -i \ket{1_{q}, 0_{m}}$.
	
	To recover the initial state $\ket{\Psi_{0}} = \ket{1_{q}, 0_{m}} \equiv (0,0,1)$ from $\ket{\Psi_{T}}$, Eq.~\eqref{eq:PsiT}, there are again several possibilities.
	Choosing again a rotation axis lying on the $xz$-plane, we perform another $\pi$-pulse around the axis $\vec n_{R}$ lying exactly halfway between 
	$\ket{\Psi_{T}}$ and the $z$-axis on the $xz$-plane. For this choice we find $\vec n_{R} = \vec n_{\perp}$ and thus 
	$\hat U_{R} = \hat U_{\perp}$.
	
	Fig.~\ref{fig:BlochSphereEnd} gives an illustration of the effect of the full sequence for a detunig between qubit and memory of $\delta\omega = - v_{\perp}$ and 
	a total free evolution time of $T = 2 \pi/\omega_{0}$. The time evolution starts and ends in the state $\ket{1_{q}, 0_{m}}$. 
	The periods of free evolution depicted by red arrows are interspersed by application of the pulses $\hat U_{\perp}$ and $\hat U_{R}$, shown in blue.
	
	The pulses $\hat U_{\perp}$ and $U_{R}$, given by Eq.~\eqref{eq:UEcho}, act in general on both qubit and memory and might be hard to realize experimentally.  
	However, our choice of rotation axis on the $xz$-plane for both pulses enables us to give a protocol for the $T_{1}$-echo sequence 
	which does not require direct unitary control of either qubit or memory. The whole sequence can in fact be realized utilizing only Hamiltonian evolution.
	In this scenario, the pulses $\hat U_{\perp}$ and $\hat U_{R}$ correspond to periods during which the qubit is tuned to a different working point than for the free 
	evolution time $T$.
	The working point for these sequences is easy to calculate, when remembering that the pulse we want to effect is a $\pi$-pulse around an axis which is perpendicular
	to the original free  evolution axis $n_{0}$. The position of $n_{0}$ is completely governed by the detuning angle $\xi_{0} = \arctan{(v_{\perp}/ \delta\omega)}$.
	Therefore, to have the system evolve around an axis perpendicular to $n_{0}$, we simply have to change the detuning $\delta\omega$ such, that the new
	detuning angle $\xi_{1}$ encloses the angle $\pi/2$ with $\xi_{0}$. We find
	\begin{equation}
		\delta\omega_{1} = - v_{\perp}^{2} / \delta\omega \,.
	\end{equation}
	In order to generate a rotation angle of $\pi$ we only have to leave the qubit at this detuning for a time $t_{\pi} = \pi / \omega_{1}$, where 
	$\omega_{1} = \sqrt{v_{\perp}^{2} + \delta\omega_{1}^{2} }$ is the oscillation frequency at the new bias point.
	For these parameters of detuning and evolution time, we find $\hat U_{1}(t_{\pi}, \xi_{1}) = \hat U_{\perp}(\xi_{0})$.
	
	Fig.~\ref{fig:PulseSequenceDetuned} shows an illustration of the $T_{1}$-echo sequence using only Hamiltonian evolution.
	The red line represents the bias point of the qubit as a function of time. Resonance with the memory is indicated by the dashed black line.
	
	We want to stress that the minimum timescale for the application of the $T_{1}$-echo sequence is given by $2 t_{\pi} = 2\pi/\omega_{1}$. 
	In contrast the minimum timescale for a full SWAP operation (transferring information into a memory and back again) is $2 t_{SWAP} = 2\pi / v_{\perp}$,
	which is always longer or at minimum the same length as $2 t_{\pi}$.
	
\section*{Effects of decoherence}

	\begin{figure}[htbp]
		\begin{center}
			\includegraphics[width=.9\columnwidth]{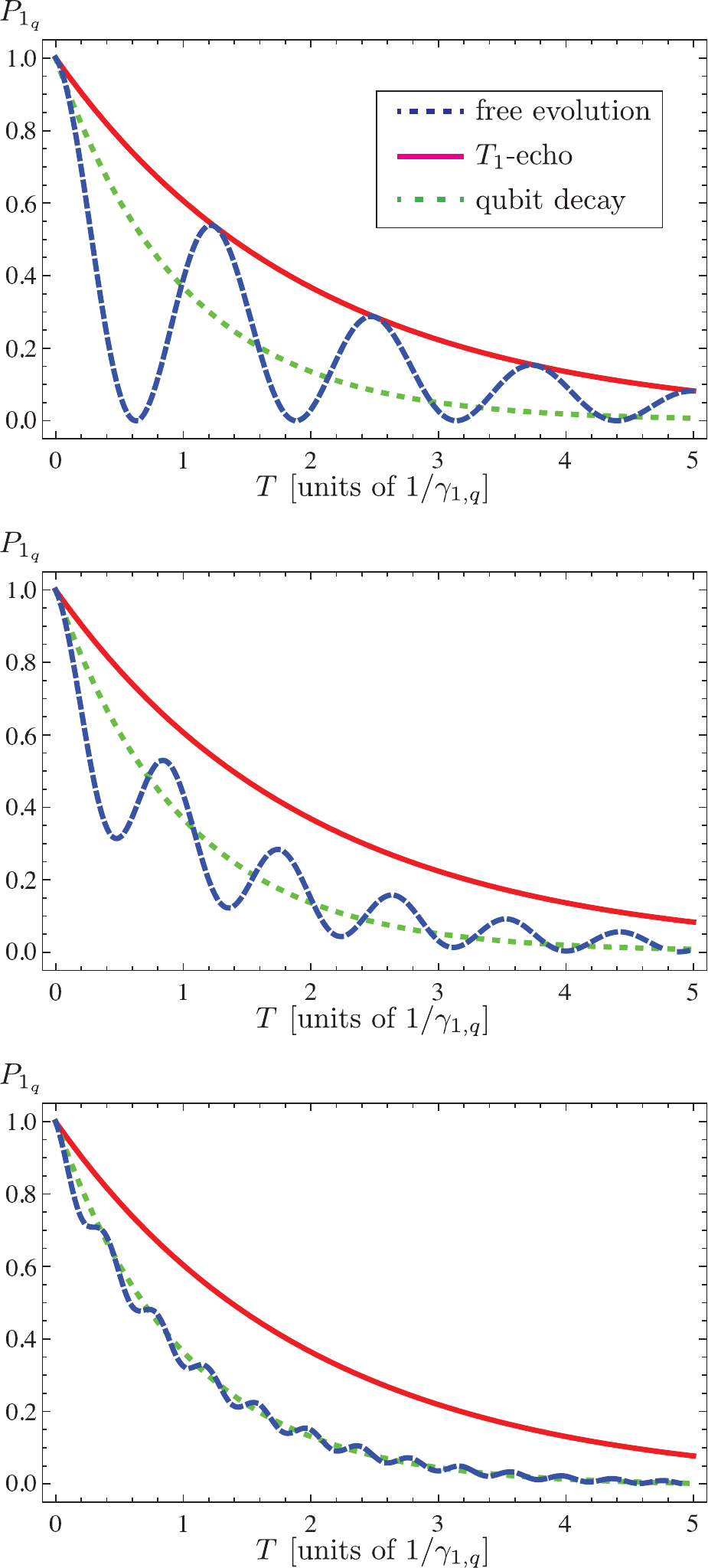}
			\caption{(Color online) Time evolution of the qubits excited state population $P_{1_{q}}$ as a function of free evolution time $T$ with (red solid) and without 
				(blue dashed) application of the $T_{1}$-echo sequence and for three different detunings between qubit and memory. 
				The dotted green line shows the decay of the qubit itself without any interactions with other quantum systems.
				The pulses are assumed to be infinitely sharp in time and there is relaxation acting on only the qubit, not the memory.
				Parameters are (in units of $\gamma_{1,q}$): $\gamma_{1,m} = \gamma_{\varphi,q} = \gamma_{\varphi,m} = 0$, 
				$v_{\perp} = 5$, $\delta\omega = 0 \text{ (top), } 5 \text{ (middle), } 15 \text{ (bottom)}$
			}
			\label{fig:T1EchoPulse}
		\end{center}
	\end{figure}
	\begin{figure}[htbp]
		\begin{center}
			\includegraphics[width=.9\columnwidth]{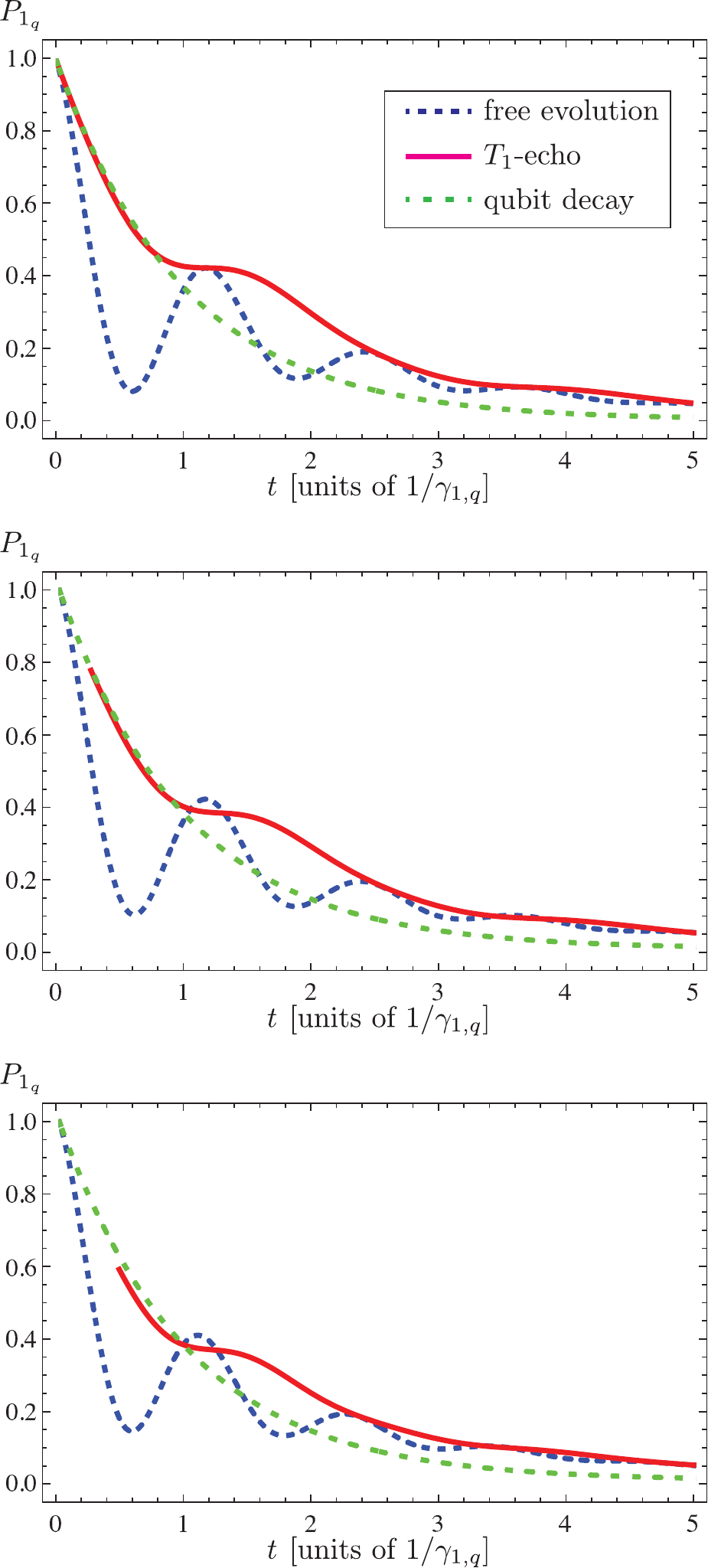}
			\caption{(Color online) Time evolution of the qubits excited state population $P_{1_{q}}$ as a function of time with (solid red) and without (dashed blue) 
				application of the $T_{1}$-echo sequence and for three different detunings between qubit and memory. 
				The pulses are realized in this case using detuned Hamiltonian evolution, see text.
				The green dotted line shows again simple exponential decay with the rate $\gamma_{1,q}$.
				The red line does not start at the origin due to the finite pulse time $2 t_{\pi}$ needed for application of the sequence.
				Parameters are (in units of $\gamma_{1,q}$): $\gamma_{\varphi, q} = 0.5$, $\gamma_{1,m} = \gamma_{\varphi,m} = 0$, 
				$v_{\perp} = 5$, $\delta\omega = 0 \text{ (top), } 1 \text{ (middle), } 2 \text{ (bottom)}$.
			}
			\label{fig:T1EchoH}
		\end{center}
	\end{figure}
	
	\begin{figure}[htbp]
		\begin{center}
			\includegraphics[width=.9\columnwidth]{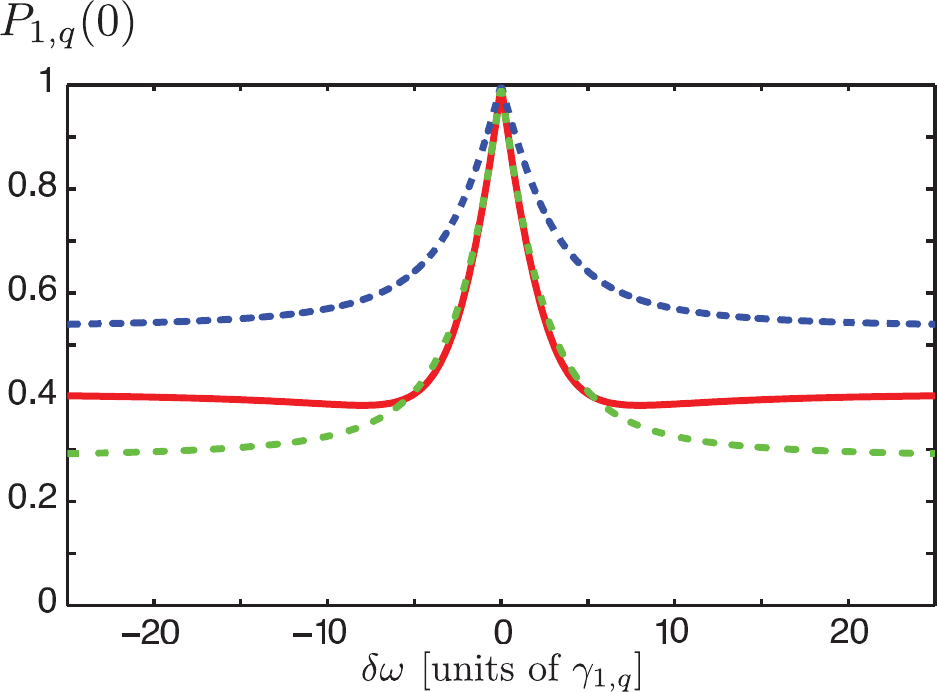}
			\caption{(Color online) Initial decay of the qubit state vector due to finite pulse length $t_{\pi}$ obtained from numerical calculations. 
				The red solid curve shows the probability to find the qubit in its excited state as a function of detuning $\delta\omega$, 
				after only the pulses have been applied to the system via detuned Hamiltonian evolution. 
				The total time that has passed is thus $2 t_{\pi}$, while no free evolution has taken place, so the free evolution time $T$ is zero.
				The green dotted curve for comparison shows 
				the value of $e^{- \gamma_{1,q} 2 t_{\pi}}$, meaning the value of qubit excitation probability after application of the pulses without any influence of the TLS.
				The blue dashed line finally shows decay with the composite rate $\gamma_{+}$.
				One can see that for small initial detuning, the pulses lead to decay dominated by the qubit rate while for stronger initial detuning it crosses over to 
				decay of the coupled systems.
				Parameter values are (in units of $\gamma_{1,q}$): $\gamma_{\varphi,q} = 0.5$, $\gamma_{1,m} = \gamma_{\varphi,m} = 0$, $v_{\perp} = 5$. 
			}
			\label{fig:T1HInit}
		\end{center}
	\end{figure}

	Up to now we considered the ideal case of purely unitary evolution governed by the Hamiltonian Eq.~\eqref{eq:H0}.
	In a real-world scenario, however, the time-evolution will not only be affected by the coupling of the two systems to each other, 
	but also by their inevitable coupling to their respective environments. 
	To show the workings of the $T_{1}$-echo sequence, we therefore have to take into account the effects of decoherence on the time-evolution. 
	As stated initially, the sequence has the properties, that it might serve to protect the qubit state against relaxation in the specific case where the additional coupled system is a memory, 
	i.e. its coherence properties are better than those of the qubit. 
	We will therefore focus on this situation in all our following illustrations although our theoretical treatment is valid in the general case.
	
	To obtain the time-evolution including decoherence, we solve a Bloch-Redfield master equation~\cite{Bloch:1957, Redfield:1957}
	 in secular approximation~\cite{Muller:2009} for the density matrix $\rho(t)$ of the coupled system of qubit and memory.
	We introduce individual relaxation rates for qubit and memory as $\gamma_{1,q/m}$. Similarly we write the pure dephasing rates as $\gamma_{\varphi, q/m}$
	with the total dephasing rate given by $\gamma_{2} = 1/2 \gamma_{1} + \gamma_{\varphi}$.
	
	For the straightforward case, where the pulses $\hat U_{\perp}$ and $\hat U_{R}$ are applied as simple unitary gates with no time cost, 
	we are able to give simple and intuitive analytical results as well as an analytical solution of full process tomography, 
	demonstrating the high degree of protection $T_{1}$-echo offers against energy relaxation.
	For the more realistic case where the pulses are effected employing purely Hamiltonian evolution, as illustrated in Fig.~\ref{fig:PulseSequenceDetuned}, 
	we will present numerical results and show that also in this case an effective protection from decoherence can be achieved.
	
	For clarity we will first focus on the case of only energy relaxation acting on the qubit and neglect pure dephasing.
	Then, if the pulses are infinitely strong, i.e., it does not take any time to apply them, 
	we arrive at a simple solution for the density matrix $\rho(t)$ of the coupled system.
	For a starting state of $\ket{1_{q}, 0_{m}}$ we calculate the population of the qubits excited state $P_{1,q} = \frac{1}{2} \left( \text{tr}\{ \rho(T) \sigma_{z} \} + 1 \right)$ 
	as a function of the free evolution time $T$ and get
	\begin{equation}
		P_{1,q} (T) = e^{-\gamma_{+} T} \,,
		\label{eq:P1q}
	\end{equation}
	where $\gamma_{+} = \frac{1}{2} \left( \gamma_{1,q} + \gamma_{1,m} \right)$ is the mean of the individual relaxation rates of qubit and memory.
	Notably this result is independent of detuning $\delta\omega$ between qubit and memory.
	
	Fig.~\ref{fig:T1EchoPulse} shows the time evolution of the qubit's excited state population as a function of free evolution time $T$ from analytical calculations
	when considering only relaxation acting on both subsystems.
	The blue dashed lines show free evolution without the application of any pulses. 
	The red solid lines give the results after one subjects the system to the $T_{1}$-echo pulses.
	With the sequence, independent of $\delta\omega$, a simple exponential decay is seen, following Eq.~\eqref{eq:P1q}.
	As a comparison, the green dotted lines shows the exponential decay $\propto e^{-\gamma_{1,q} t }$ of the qubit itself, without any coupling to another quantum system.
	For zero detuning (top panel), the free evolution without any pulses shows strong coherent oscillations of the excitation between qubit and memory.
	After  application of a $T_{1}$-echo pulse, the resulting decay is given by the envelope of the coherent decay curve (cf. Ref.~\cite{Muller:2009}).
	For non-zero detuning (lower two panels), the amplitude of the oscillations decreases and its frequency $\omega_{0}$ increases. 
	The effective decay changes from decay of the mixed systems for small detuning to more qubit-like behavior for stronger detuning. 
	In contrast, the decay of the qubit state after application of the $T_{1}$-echo sequence stays the same and is given always by Eq.~\eqref{eq:P1q}.

	The whole sequence including decoherence can be thought of as a quantum dynamical map $\varepsilon(\rho_{q})$ acting on the density matrix $\rho_{q}$ of the qubit. 
	We can write this map in terms of a basis set $\left\{ E_{n} \right\}$ of all operators acting on the Hilbert space of the qubit as
	\begin{equation}
		\varepsilon (\rho_{q}) = \sum_{mn} \chi_{mn} E_{m} \rho_{q} E_{n}^{\dagger}
	\end{equation}
	with the coefficient Matrix $\chi$. 
	Choosing as basis set $\left\{ E_{n} \right\} = \left\{ \mathbb{1}, \sigma_{x}, \sigma_{y}, \sigma_{z} \right\}$ we get for the matrix $\chi$
	\begin{widetext}
		\begin{equation}
			\chi = \frac{1}{4} \left( \begin{array}{cccc}
				1 + e^{- \gamma_{+} t} - 2 e^{-\frac{1}{2} \gamma_{+} t} \cos{\epsilon t } & 0 & 0 & 1 - e^{- \gamma_{+} t} - 2 i e^{-\frac{1}{2} \gamma_{+} t} \sin{\epsilon t} \\
				0 & 1 - e^{- \gamma_{+} t} & -i \left( 1 - e^{- \gamma_{+} t} \right) & 0 \\
				0 & i \left( 1 - e^{- \gamma_{+} t} \right) & 1 - e^{- \gamma_{+} t} & 0 \\
				1 - e^{- \gamma_{+} t} + 2 i e^{-\frac{1}{2} \gamma_{+} t} \sin{\epsilon t} & 0 & 0 &  1 + e^{- \gamma_{+} t} + 2 e^{-\frac{1}{2} \gamma_{+} t} \cos{\epsilon t }
			\end{array} \right) \,,
		\end{equation}
	\end{widetext}
	where the oscillation frequency is $\epsilon \approx \epsilon_{q} \approx \epsilon_{m}$ and the sole appearing decay rate is
	$\gamma_{+}$. Again, this result is valid for only relaxation acting on both subsystems.
	In this case the only relevant decay rate in the time evolution is the mean of the individual decay rates of the two subsystems, 
	as can also be seen in Fig.~\ref{fig:T1EchoPulse}.
		
	The results presented up to now are for the ideal case when perfect pulses are available and no pure dephasing acts on either one of the subsystems. 
	Above we have given the possibility to generate the $T_{1}$-echo sequence using only detuned free evolution without the need for high-frequency controls.
	In this proposal however, the pulses will take a finite amount of time $t_{\pi}$ during which decoherence will act on both systems and therefore the results will be worse
	than for the case of infinitely short pulses.
	Additionally, the above calculations did not include pure dephasing as a source of decoherence.
	
	Fig.~\ref{fig:T1EchoH} shows the time evolution of the qubits excited state population $P_{1,q}$ as a function of total time $t$ and for different detunings
	when the sequence pulses are realized by detuned Hamiltonian evolution.
	These results are from numerical calculations solving a Lindblad master equation for the time evolution of the systems density matrix.
	Both relaxation rates $\gamma_{1}$ as well as pure dephasing rates $\gamma_{\varphi}$ have been included in the calculations.
	The red solid line again shows the time evolution of the qubits excited state population $P_{1,q}$ after application of the $T_{1}$-echo, 
	while the dashed blue shows the free evolution without any pulses.
	The green dotted line is again for comparison with simple qubit decay $\propto e^{-\gamma_{1,q} t}$.
	In this situation the qubit state population after applying the pulses still shows some residual oscillations, due to the mixing of qubit and memory states during the pulses.
	The overall decay is somewhat faster than predicted by Eq.~\eqref{eq:P1q}, mainly due to the additional pure dephasing acting on the system. 
	With the exception of an initial residual oscillation, which is suppressed on a timescale given by the dephasing rates $\gamma_{2}$, the probability of finding
	the qubit in its excited state $P_{1,q}$ is a smooth function of time, highlighting the performance of $T_{1}$-echo.
	Additionally, in the illustrated case of interaction with a memory element, the lifetime of the qubit state is enhanced as compared to the free case.
	
	To quantify the loss of qubit coherence during the application of the echo pulses, 
	Fig.~\ref{fig:T1HInit} shows a plot of the initial probability of finding the qubit in the excited state, $P_{1,q}(0)$, when no free evolution time $T$ has passed, 
	i.e. only the pulses have been applied via detuned evolution.
	The solid red line shows $P_{1,q}(0)$ as  a function of initial detuning $\delta\omega$ including both relaxation and dephasing on the qubit. 
	For comparison, the dotted green line gives the excitation probability of the uncoupled qubit after the time $2 t_{\pi}$ it takes to perform the gates. 
	Finally, the dashed blue line, gives the value of $e^{-\gamma_{+} 2 t_{\pi}}$, i.e., the effective decay of the resonantly coupled system. 
	As one can see, close to resonance, where for zero time $T$ the excitation resides mostly in the qubit during the pulses, the decay during detuned evolution is 
	\emph{qubit-like}, i.e. $\propto e^{-\gamma_{1,q} 2 t_{\pi}}$. For stronger initial detuning however, the detuning during application of the gates, $\delta\omega_{1}$,
	is close to resonant and therefore the effective decay becomes similar to the decay of the resonantly coupled systems when including dephasing~\cite{Muller:2009}.

\section*{Discussion}

	It is instructive to consider the limits of the proposed pulses.
	In one limit, when qubit and memory are initially in resonance, $\delta\omega= 0$, the pulses correspond to a very strong detuning for an equally short time. 
	This effectively leads to a phase gate acting on the system, 
	which in the effective Bloch-sphere picture transfers the state from one hemisphere to the opposite one (cf. Fig.~\ref{fig:BlochSphereInit}).
	In the other limit, strong initial detuning, $\delta\omega \gg v_{\perp}$, we find $\delta\omega_{1} = 0$, 
	In this case the pulses $U_{\perp}$ and $U_{R}$ both correspond to a full transfer of quantum information between the qubit and the memory, 
	and thus to a full SWAP between qubit and memory. 	
	This situation provides an intuitive explanation for the resulting effective relaxation rate $\gamma_{+}$ in Eq.~\eqref{eq:P1q}: 
	the excitation first rests in the qubit for a time $T/2$ before being transferred into the memory and resting there for another timespan $T/2$. 
	The effective decay is therefore governed by the average of the two individual rates. 
	In the case of strong detuning, therefore, using a traditional SWAP sequence, where the information resides for the full time $T$ in the memory is more advantageous. 
	In the opposite case, however, the $T_{1}$-echo not only negates the effects of coherent coupling 
	but also provides a very fast way of refocusing the qubit state and offering additional protection from relaxation.
	
	The treatment of dephasing above was done by introducing additional pure dephasing rates $\gamma_{\varphi}$ into the Redfield-tensor.
	A more realistic treatment has to take into account the origin of pure dephasing, that is slow fluctuations in the level energy of qubit and memory. 
	The effect of these fluctuations on the evolution is a change of the detuning $\delta\omega$ and thus of both the angle $\xi_{0}$ as well as 
	the oscillation frequency $\omega_{0}$.
	As long as the evolution time $T$ is shorter than the inverse frequency of the fluctuation, 
	the $T_{1}$-echo sequence will refocus the state of the qubit also in this case, in exact analogy to Hahn-echo. 
	For the special case of zero detuning between qubit and memory, $\delta\omega = 0$, this has already been experimentally demonstrated in Ref.~\cite{Gustavsson:2012}.
	In that work it was shown that additional echo pulses $\hat U_{\perp}$ at equally spaced time intervals can provide additional protection from dephasing~\cite{Bylander:2011}. 
	This conclusion however is only true for the slow noise spectrum leading to pure dephasing. Concatenation of pulses will then lead to a further decreased sensitivity of the 
	system to the low-frequency parts of the spectrum.
	Considering energy relaxation into a Markovian environment, on the other hand, Eq.~\eqref{eq:P1q} describes the maximal achievable protection with our scheme.

	The realization of the $T_{1}$-echo we have focused our analyzis on here, using detuned Hamiltonian evolution, is only one of many possibilities.
	In concrete schemes more specific sequences might be more practical.
	Depending on the implementation of qubit and memory, two qubit gates might be readily available, and depending on specific gate times might outperform 
	the implementation without high-frequency controls we  showed here. 
	
\section*{Conclusions}

	We presented the $T_{1}$-echo sequence as a method to dynamically protect the state of a qubit from the effects of interaction with another coherent quantum system.
	In the case where the additional system shows better coherence properties than the qubit itself, it additionally serves to improve
	the qubit relaxation time. 
	The sequence is formally similar to the well known Spin-Echo sequence and can be experimentally realized 
	without the need for high-frequency controls.
	Implementations in different systems are possible, the only prerequisite for the proposed implementation is the tune-ability of the qubit transition frequency.
	Possible realizations are in superconducting qubit systems, 
	where either intrinsic two-level defects~\cite{Neeley:2008} or superconducting resonators~\cite{Mariantoni:2011} would play the role of the additional elements, 
	and implementation is straightforward.
	
\section*{Acknowledgements}
	
	We gratefully acknowledge discussion with D. Pappas. M. Sandberg and M. Vissers and want to thank A. Blais, G. Grabovskij and J. Lisenfeld
	for their comments on the manuscript.
	We acknowledge financial support by CFN of DFG as well as NSERC.

\bibliography{T1Echo}

\end{document}